\documentclass[fleqn,twoside]{article}
\usepackage{myespcrc2}
\usepackage{axodraw}

\usepackage{epsfig}
\usepackage{graphicx}
\usepackage{dcolumn}
\usepackage{bm}

\newcommand{\be}{\begin{equation}}
\newcommand{\ee}{\end{equation}}
\newcommand{\nn}{\nonumber}
\newcommand{\bea}{\begin{eqnarray}}
\newcommand{\eea}{\end{eqnarray}}
\newcommand{\bfig}{\begin{figure}}
\newcommand{\efig}{\end{figure}}
\newcommand{\bc}{\begin{center}}
\newcommand{\ec}{\end{center}}

\def\spa#1.#2{\langle#1\,#2\rangle}
\def\spb#1.#2{[#1\,#2]}
\def\spab#1.#2.#3{\langle\mskip-1mu{#1} 
                  | #2 | {#3}]}

\def\spba#1.#2.#3{[\mskip-1mu{#1} 
                  | #2 | {#3}\rangle}

\def\spbb#1.#2.#3.#4{[\mskip-1mu{#1} 
                     | {#2}  {#3} | {#4}]}

\def\spaa#1.#2.#3.#4{\langle\mskip-1mu{#1} 
                     | {#2}  {#3} | {#4}\rangle}

\def\spaab#1.#2.#3.#4.#5{\langle\mskip-1mu{#1} 
                     | {#2} \ {#3} \ {#4} | {#5}]}

\def\es#1.#2{s_{#1 #2}}

\begin{document}

\begin{titlepage}

\title{Six-Photon Amplitudes}

\author{T.\ Binoth, and G.\ Heinrich, \\
{\it School of Physics, The University of Edinburgh, Edinburgh EH9 3JZ, Scotland} 
\vskip 0.5cm
        T.\ Gehrmann, and P.\ Mastrolia \\
{\it Institut f\"ur Theoretische Physik, Universit\"at Z\"urich, CH-8057 Z\"urich, Switzerland}
}


\begin{abstract}
We present analytical results for all six-photon
helicity amplitudes. For the computation of this
loop induced process two recently developed methods,
based on form factor decomposition and on multiple
cuts, have been used.
We obtain compact results, demonstrating
the applicability of both methods to one-loop amplitudes relevant to
precision collider phenomenology.
\end{abstract}

\maketitle
\end{titlepage}

\thispagestyle{myheadings}
\markright{{\tt Edinburgh 2007/7, ZU-TH 08/07}}


The self-interaction of photons (light-by-light scattering) 
mediated through a virtual charged fermion loop 
is a fundamental,
although yet unobserved, prediction of quantum 
electrodynamics. The corresponding multi-photon 
scattering amplitudes are of outstanding theoretical interest, since 
they exhibit a high degree of symmetry.  They can be used to establish and 
further develop methods for the calculation of virtual corrections to 
multi-leg processes, and to study symmetry patterns in the results, thus 
providing further insight into the analytical structure of quantum field 
theories at the loop level. 

In the past, the four-photon amplitudes were derived at  one loop 
for massless and massive fermions~\cite{karplus}, 
and at two loops for massless 
fermions~\cite{gamtwo}.
At two loops, four-photon scattering 
in supersymmetric Yang-Mills theories~\cite{gamsusy}
was studied as well,
yielding first evidence for a leading transcendentality behaviour, which 
was uncovered subsequently also in multi-gluon 
amplitudes~\cite{transc}, and has sparked 
many new developments (see~\cite{transc2} and references therein)
in the study of multi-loop scattering  amplitudes in 
super-Yang-Mills theories and perturbative gravity. 

The computation of one-loop multi-particle amplitudes is currently 
among the most pressing issues in the preparation of precision 
next-to-leading order (NLO) calculations 
for the upcoming CERN LHC experiments. Given the large variety of potentially 
interesting multi-particle final states, automated methods for 
one-loop corrections would be very desirable, and are currently under 
intense 
development~\cite{method1,method2,method3,method4,Ferroglia:2002mz,Kurihara:2005ja,method6,Czakon:2005rk,Britto:2004nc,Britto:2005ha,Britto:2006sj,Mastrolia:2006ki,method7,method7bis,method8,method9,method10,nagy,sixglu,sixfer,Lazopoulos:2007ix,Anastasiou:2007qb}.
These methods range from purely analytical schemes to completely numerical 
approaches.

Compact analytical expressions for amplitudes involving $n>4$ external 
photons were obtained only for specific helicity configurations. All 
amplitudes with odd $n$ vanish due to parity 
conservation; amplitudes with even $n>4$ vanish if all or all but one 
photons have the same helicity~\cite{mahlon1}.  
For $n=6$ external 
photons, one finds therefore only two non-vanishing amplitudes, which 
we denote by $A_6(--++++)$ and $A_6(-+-+-+)$. An analytical 
expression for  $A_6(--++++)$ was computed already long 
ago~\cite{mahlon2}, using the 
method described in~\cite{mahlon1}. $A_6(-+-+-+)$ was obtained recently 
using a purely numerical method for the loop integration. In this paper, 
we use two completely different recently developed methods 
(based either on form factor decomposition~\cite{method1} 
or on multiple 
cuts~\cite{Britto:2004nc,Britto:2005ha,Britto:2006sj,Mastrolia:2006ki}) 
to compute 
$A_6(--++++)$ and $A_6(-+-+-+)$, obtaining compact results which respect 
the symmetry properties of the process under consideration. Besides allowing 
a prediction for the process $\gamma\gamma \to 4 \gamma$, 
our results serve 
as a highly non-trivial proof of applicability of both methods used, and 
illustrate how form factor-based and cut-based techniques can be matched 
onto each other in detail.

\section{Structure of the Amplitudes}\label{sec:structure}

Despite the absence of a corresponding tree-level
process and the (Feynman) diagram-by-diagram UV-finiteness
in four dimensions, the cut-constructibility
of the 6-photon amplitudes is not guaranteed,
in accordance with the power counting argument in \cite{Bern:1994zx}.
The fact that the rational parts of the six-photon amplitudes actually
do evaluate to zero was shown in \cite{binoth2}.
Consequently, the amplitude can be written
as a linear combination of poly-logarithms and transcendental
constants, associated to a known
basis of functions, the master integrals (MI),
formed by box-, triangle- and bubble-type integrals
\cite{tHooft:1979,Bern:1993kr,Binoth:2001vm}.

The result for the amplitude $A_6(--++++)$ has the following structure 
\bea
&&A_6(--++++) = \frac{e^6}{(4\pi)^2} 
\sum\limits_{\sigma \in \mathcal{S}_4/(Z_2\times Z_2)} \Big[
\nonumber\\ &&
\qquad 
  {\cal F}_1(s_{1\sigma_3},s_{1\sigma_4},s_{1\sigma_3\sigma_4})
+ {\cal F}_1(s_{2\sigma_3},s_{2\sigma_4},s_{2\sigma_3\sigma_4})\nonumber\\ &&
\qquad - {\cal F}_{2B}(s_{1\sigma_3\sigma_5},s_{1\sigma_4\sigma_5},s_{1\sigma_5},s_{2\sigma_6})\nonumber\\ &&
\qquad - {\cal F}_{2B}(s_{2\sigma_3\sigma_5},s_{2\sigma_4\sigma_5},s_{2\sigma_5},s_{1\sigma_6})
\Big] \ ,
\label{eq:twominusdeco}
\eea
where
\bea
 {\cal F}_i = d_i \times F_i \qquad (i=1, 2B) \;.\nn
\eea
$F_1,F_{2B}$ are the finite parts of the {\it one-mass} 
and {\it two-mass-easy} box functions \cite{Bern:1993kr,Binoth:2001vm}, and 
 $d_1$ and $d_{2B}$ are their coefficients, 
that can be obtained from (\ref{eq:twominusD1}) and (\ref{eq:twominusD2B}) 
below.
The sum in Eq.(\ref{eq:twominusdeco}) runs over the discrete quotient group
$\mathcal{S}_4(3,4,5,6) / (Z_2(3,4)\times Z_2(5,6))$ which has 6 elements. 
The quotient space structure
can be infered from the symmetry properties of the combination
of basis functions. It is invariant under  interchanging 
the indices $\sigma_3 \leftrightarrow \sigma_4$ and 
$\sigma_5 \leftrightarrow  \sigma_6$. The permutations generate exactly those functions which 
 are allowed by the cutting rules.

The result for the amplitude $A_6(-+-+-+)$ has the following structure 
\bea
&&A_6(-+-+-+) = \frac{e^6}{(4\pi)^2} \Big[\nonumber \\&&
\sum\limits_{(\sigma,\tau) \in \mathcal{S}_3/Z_2 \times \mathcal{S}_3/Z_2} 
 {\cal F}_1(s_{\sigma_1\tau_2},s_{\tau_2\sigma_3},s_{\tau_4\sigma_5\tau_6})\nonumber\\&&
+
\sum\limits_{(\sigma,\tau) \in \mathcal{S}_3/Z_2 \times \mathcal{S}_3/Z_2} 
 {\cal F}_1(s_{\tau_6\sigma_1},s_{\sigma_1\tau_2},s_{\sigma_3\tau_4\sigma_5})\nonumber\\&&
+
\sum\limits_{(\sigma,\tau) \in \mathcal{S}_3 \times \mathcal{S}_3} 
 {\cal F}_{2A}(s_{\sigma_1\tau_2},s_{\tau_2\sigma_3\tau_4},s_{\sigma_3\tau_4},s_{\sigma_5\tau_6})\nonumber\\&&
+
\sum\limits_{\tau \in \mathcal{S}_3} 
 {\cal I}_{3}(s_{1\tau_2},s_{3\tau_4},s_{5\tau_6})
\Big] \ ,
\label{eq:threeminusdeco}
\eea
where
\bea
 {\cal F}_i = d_i \times F_i \quad (i=1, 2A) \ , {\rm \ and \ } 
{\cal I}_3 = c_3 \times I_3 \;.\nn
\eea
$F_1,F_{2A}$ are the finite parts of the {\it one-mass} 
and {\it two-mass-hard} box functions, $I_3$ is the triangle function with 
3 off-shell legs and 
$d_1$, $d_{2A}$, and $c_3$ are obtained from 
(\ref{eq:threeminusD1}), (\ref{eq:threeminusD2A}) and (\ref{eq:threeminusC3}) 
given in section \ref{sec:cut}.
The different permutation groups 
in (\ref{eq:threeminusdeco}), 
i.e. $\mathcal{S}_3(1,3,5)/Z_2(1,3) \times \mathcal{S}_3(2,4,6)/Z_2(4,6)$
for the first term,
$\mathcal{S}_3(1,3,5)/Z_2(3,5) \times \mathcal{S}_3(2,4,6)/Z_2(2,6)$
for the second term,
$\mathcal{S}_3(1,3,5) \times \mathcal{S}_3(2,4,6)$
for the third one and $\mathcal{S}_3(2,4,6)$ for the last term, 
contain 9, 9, 36 and 6 ele\-ments respectively.
They generate all possible different basis functions $F_1$, $F_{2A}$
and $I_{3}$ which are allowed by cutting rules.
Note that each cut is corresponding to a certain Madelstam variable.
Compatibility with cutting rules means that no Mandelstam variables
$s_{ij}$ or $s_{ijk}$ appear as a function argument such that 
the corrsponding legs $i,j,k$ have like-sign helicities. 
This is certainly only true for amplitudes with massless particles.

\section{Cut-Construction}\label{sec:cut}

As the six-photon amplitudes are cut-constructible,  
the computational effort is  reduced to the computation of 
the rational coefficients of a linear combination of 
master integrals as described in section \ref{sec:structure}.
According to the principle of unitarity-based methods \cite{Bern:1994zx}, 
the exploitation of the unitarity-cuts of each master integral enables the extraction
of the corresponding coefficient from the amplitude.
To that aim, we employ 
the quadruple-cut technique~\cite{Britto:2004nc} for box-coefficients, 
the triple-cut integration~\cite{Mastrolia:2006ki} for triangle-coefficients, 
and the double-cut integration~\cite{Britto:2005ha,Britto:2006sj} 
for bubble-coefficients, by sewing in the multiple cuts the 
QED tree-level amplitudes given in \cite{Ozeren:2005mp}.
Spinor algebra and numerical evaluation of spinor products has been implemented
in a {\tt Mathematica} package \cite{Sp4M}.
In the following we use by now standard spinor notation, 
with $\spab a.P_{i \ldots j}.b = \langle a^-| P_{i \ldots j}| b^-\rangle$, 
and multi-particle
momenta defined as $P_{i \ldots j} = p_i + \ldots + p_j$, with $p_k$ being the 
momentum of the $k$-th external photon, considered incoming.


\markright{}
\subsection{Construction of $A_6(- - + + + +)$}

The analytic expression for $A_6(- - + + + +)$ was computed 
by Mahlon \cite{mahlon2}.
It can be expressed in terms of the two classes of box functions 
$F_1$ and $F_{2B}$ as shown in eq.~(\ref{eq:twominusdeco}).
Using the symmetry of the amplitude, 
it is sufficient to compute the coefficients
of a representative box-function for each of the two classes 
(and their parity-conjugates).
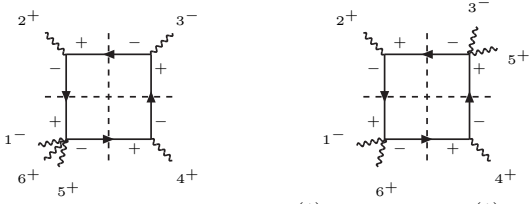
\begin{figure}[h]
\begin{center}
\vskip 0.5cm
\begin{picture}(0,0)(0,0)
\SetScale{0.8}
\SetWidth{0.8}
\ArrowLine(-20,-20)(20,-20)
\ArrowLine(20,-20)(20,20)
\ArrowLine(20,20)(-20,20)
\ArrowLine(-20,20)(-20,-20)
\Photon(20,-20)(30,-30){1}{4}   \Text( 30,-30)[]{{\tiny{$4^+$}}}
\Photon(20,20)(30,30){1}{4}     \Text( 30,30)[]{{\tiny{$3^-$}}} 
\Photon(-20,20)(-30,30 ){1}{4}  \Text(-30,30)[]{{\tiny{$2^+$}}}
\Photon(-20,-20)(-33,-23){1}{4} \Text(-35,-15)[]{{\tiny{$1^-$}}}
\Photon(-20,-20)(-30,-30){1}{4} \Text(-30,-30)[]{{\tiny{$6^+$}}}
\Photon(-20,-20)(-23,-33){1}{4} \Text(-15,-35)[]{{\tiny{$5^+$}}}
\DashLine(0,30)(0,-30){3}  
%
\Text(-10,+20)[]{{\tiny{$^+$}}}  \Text(+10,20)[]{{\tiny{$^-$}}}
\Text(-10,-20)[]{{\tiny{$^-$}}} \Text(+10,-20)[]{{\tiny{$^+$}}}
\DashLine(-30,0)(30,0){3}  
%
\Text(-20, 10)[]{{\tiny{$^-$}}}  \Text(+20, 10)[]{{\tiny{$^+$}}}
\Text(-20,-10)[]{{\tiny{$^+$}}}  \Text(+20,-10)[]{{\tiny{$^-$}}}
\end{picture}
\hspace*{4.0cm}
\begin{picture}(0,0)(0,0)
\SetScale{0.8}
\SetWidth{0.8}
\ArrowLine(-20,-20)(20,-20)
\ArrowLine(20,-20)(20,20)
\ArrowLine(20,20)(-20,20)
\ArrowLine(-20,20)(-20,-20)
\Photon(20,-20)(30,-30){1}{4}   \Text( 30,-30)[]{{\tiny{$4^+$}}}
\Photon(20,20)(33,23){1}{4}     \Text( 35,15)[]{{\tiny{$5^+$}}} 
\Photon(20,20)(23,33){1}{4}     \Text( 20,35)[]{{\tiny{$3^-$}}}
\Photon(-20,20)(-30,30){1}{4} \Text(-30,30)[]{{\tiny{$2^+$}}}
\Photon(-20,-20)(-33,-23){1}{4} \Text(-35,-15)[]{{\tiny{$1^-$}}}
\Photon(-20,-20)(-23,-33){1}{4} \Text(-15,-35)[]{{\tiny{$6^+$}}}
\DashLine(0,30)(0,-30){3}  
%
\Text(-10,+20)[]{{\tiny{$^+$}}}  \Text(+10,20)[]{{\tiny{$^-$}}}
\Text(-10,-20)[]{{\tiny{$^-$}}} \Text(+10,-20)[]{{\tiny{$^+$}}}
\DashLine(-30,0)(30,0){3}  
%
\Text(-20, 10)[]{{\tiny{$^-$}}}  \Text(+20, 10)[]{{\tiny{$^+$}}}
\Text(-20,-10)[]{{\tiny{$^+$}}}  \Text(+20,-10)[]{{\tiny{$^-$}}}
\end{picture}
\end{center}
\caption{
Quadruple-Cuts, $\hat{d}_1^{(1)}$ (left) and $\hat{d}_{2B}^{(1)}$ (right).
Reverse internal-helicity counterparts understood.
}
\label{FIG:twominus}
\end{figure}

\vspace*{3mm}
\noindent{ \bf  One-mass Box $ \ (5^+ 6^+ 1^-|2^+|3^-|4^+)$}

Defining the prefactor
\bea
r_1 &=& 
8 \ {
s_{23}
s_{34}
\over
\spa 2.4^2 
\spa 2.5
\spa 4.5
\spa 2.6
\spa 4.6
} \ ,
\eea
the result of the quadruple-cut $\hat{d}_{1}^{(1)}$ for the configuration
displayed in Fig.\ref{FIG:twominus} reads
\bea
\hat{d}_1^{(1)} = 
r_1 
\spa 2.3^2
\spa 1.4^2 \,.
\label{eq:twominustilD11}
\eea
By reversing the internal helicities one gets
\bea
\hat{d}_1^{(2)} =
r_1
\spa 1.2^2
\spa 3.4^2 \,.
\label{eq:twominustilD12}
\eea
\noindent
The coefficient of $F_1(s_{23}, s_{34}, s_{561})$, the finite part of the one-mass box, reads
\bea
d_1(s_{23}, s_{34}, s_{561}) = 
2 \times {(- 2) \over s_{23} s_{34}} \times
\ { \hat{d}_1^{(1)} + \hat{d}_1^{(2)} \over 2 } \,,
\label{eq:twominusD1}
\eea
where the prefactor 2 accounts for the contribution coming from
a fermion looping in  
the opposite direction; the second factor is the standard coefficient
of the finite part of the one-mass box; and the last one is the
average of the solutions of the quadruple-cut.
The same structure is understood for the forthcoming four-point coefficients.

\vspace*{3mm}

\noindent{ \bf Two-mass easy Box $ \ (6^+ 1^-|2^+|3^- 5^+|4^+)$}

The coefficient of $F_{2B}(s_{612}, s_{235}, s_{61}, s_{35})$, 
the finite part of the considered two-mass-easy box is defined as
\bea
d_{2B}(s_{612}, s_{235}, s_{61}, s_{35}) = 
\nonumber
\eea \vspace*{-0.7cm}
\bea 
&&
2 \times {(- 2) \over s_{612} s_{235} - s_{61} s_{35}} 
\times { \hat{d}_{2B}^{(1)} + \hat{d}_{2B}^{(2)} \over 2 } \,,
\label{eq:twominusD2B}
\eea
with $\hat{d}_{2B}^{(1)}$ being the quadruple-cut for the configuration
displayed in Fig.\ref{FIG:twominus}, and $\hat{d}_{2B}^{(2)}$ 
the complementary contribution 
coming from a fermion circulating around the loop in the opposite direction.

Although the quadruple-cuts $\hat{d}_{2B}^{(1)}$ and $\hat{d}_{2B}^{(2)}$ 
are respectively different from the quadruple-cuts
$\hat{d}_{1}^{(1)}$ and
$\hat{d}_{1}^{(2)}$ in Eqs.(\ref{eq:twominustilD11}, \ref{eq:twominustilD12}),
the coefficients of $F_{2B}(s_{612}, s_{235}, s_{61}, s_{35})$ in (\ref{eq:twominusD2B}) and of 
$F_1(s_{23}, s_{34}, s_{561})$ in (\ref{eq:twominusD1}) are not independent, 
and one finds the relation,
\bea
d_{2B}(s_{612}, s_{235}, s_{61}, s_{35}) = - d_1(s_{23}, s_{34}, s_{561}) \,.
\eea

\vspace*{3mm}

\subsection{Construction of $A_6(- + - + - +)$}

The analytic expression for  $A_6(- + - + - +)$ 
is the main result of this letter.
There appear three classes of functions:
one-mass and two-mass-hard box-functions $F_1$ and $F_{2A}$, 
and three-mass triangle-function $I_3$.
Bubble-functions are absent.
As before, we compute the coefficients
of a representative function for each of the three classes
(and their parity-conjugates), and obtain
the whole amplitude by summing over all non-identical permutations of 
external particles.

\begin{figure}[h]
\begin{center}
\vskip 0.5cm
\begin{picture}(0,0)(0,0)
\SetScale{0.8}
\SetWidth{0.8}
\ArrowLine(-20,-20)(20,-20)
\ArrowLine(20,-20)(20,20)
\ArrowLine(20,20)(-20,20)
\ArrowLine(-20,20)(-20,-20)
\Photon(20,-20)(30,-30){1}{4}   \Text( 30,-30)[]{{\tiny{$6^+$}}}
\Photon(20,20)(30,30){1}{4}     \Text( 30,30)[]{{\tiny{$5^-$}}} 
\Photon(-20,20)(-23,33){1}{4} \Text(-15,35)[]{{\tiny{$4^+$}}}
\Photon(-20,20)(-33,23){1}{4} \Text(-35,15)[]{{\tiny{$3^-$}}}
\Photon(-20,-20)(-33,-23){1}{4} \Text(-35,-15)[]{{\tiny{$2^+$}}}
\Photon(-20,-20)(-23,-33){1}{4} \Text(-15,-35)[]{{\tiny{$1^-$}}}
\DashLine(0,30)(0,-30){3}  
%
\Text(-10,+20)[]{{\tiny{$^+$}}}  \Text(+10,20)[]{{\tiny{$^-$}}}
\Text(-10,-20)[]{{\tiny{$^-$}}} \Text(+10,-20)[]{{\tiny{$^+$}}}
\DashLine(-30,0)(30,0){3}  
%
\Text(-20, 10)[]{{\tiny{$^-$}}}  \Text(+20, 10)[]{{\tiny{$^+$}}}
\Text(-20,-10)[]{{\tiny{$^+$}}}  \Text(+20,-10)[]{{\tiny{$^-$}}}
\end{picture}
\hspace*{2.8cm}
\begin{picture}(0,0)(0,0)
\SetScale{0.8}
\SetWidth{0.8}
\ArrowLine(-20,-20)(20,-20)
\ArrowLine(20,-20)(20,20)
\ArrowLine(20,20)(-20,20)
\ArrowLine(-20,20)(-20,-20)
\Photon(20,-20)(30,-30){1}{4}   \Text( 30,-30)[]{{\tiny{$6^+$}}}
\Photon(20,20)(30,30){1}{4}     \Text( 30,30)[]{{\tiny{$5^-$}}} 
\Photon(-20,20)(-30,30 ){1}{4}  \Text(-30,30)[]{{\tiny{$4^+$}}}
\Photon(-20,-20)(-33,-23){1}{4} \Text(-35,-15)[]{{\tiny{$3^-$}}}
\Photon(-20,-20)(-30,-30){1}{4} \Text(-30,-30)[]{{\tiny{$2^+$}}}
\Photon(-20,-20)(-23,-33){1}{4} \Text(-15,-35)[]{{\tiny{$1^-$}}}
\DashLine(0,30)(0,-30){3}  
%
\Text(-10,+20)[]{{\tiny{$^+$}}}  \Text(+10,20)[]{{\tiny{$^-$}}}
\Text(-10,-20)[]{{\tiny{$^-$}}} \Text(+10,-20)[]{{\tiny{$^+$}}}
\DashLine(-30,0)(30,0){3}  
%
\Text(-20, 10)[]{{\tiny{$^-$}}}  \Text(+20, 10)[]{{\tiny{$^+$}}}
\Text(-20,-10)[]{{\tiny{$^+$}}}  \Text(+20,-10)[]{{\tiny{$^-$}}}
\end{picture}
\hspace*{2.6cm}
\begin{picture}(0,0)(0,0)
\SetScale{0.8}
\SetWidth{0.8}
\ArrowLine(-20,-20)(20,-20)
\ArrowLine(20,-20)(0,15)
\ArrowLine(0,15)(-20,-20)
\Photon(0,15)( 8,27){1}{4} \Text(12,30)[]{{\tiny{$4^+$}}}
\Photon(0,15)(-8,27){1}{4}  \Text(-12,30)[]{{\tiny{$3^-$}}}

\Photon(20,-20)(33,-23){1}{4} \Text(35,-15)[]{{\tiny{$5^-$}}}
\Photon(20,-20)(23,-33){1}{4} \Text(15,-35)[]{{\tiny{$6^+$}}}
\Photon(-20,-20)(-33,-23){1}{4} \Text(-35,-15)[]{{\tiny{$2^+$}}}
\Photon(-20,-20)(-23,-33){1}{4} \Text(-15,-35)[]{{\tiny{$1^-$}}}
\DashLine(0,-8)(0,-30){3}  
%
\Text(-10,-20)[]{{\tiny{$^-$}}} \Text(+10,-20)[]{{\tiny{$^+$}}}
\DashLine(-25,5)(0,-8){3}  
\DashLine(0,-8)(25,5){3}  
%
\Text(-7, 8)[]{{\tiny{$^-$}}}  \Text(+7, 8)[]{{\tiny{$^+$}}}
\Text(-17,-10)[]{{\tiny{$^+$}}}  \Text(+17,-10)[]{{\tiny{$^-$}}}
\end{picture}
\end{center}
\caption{
From left to right:
quadruple-cuts, $\hat{d}_{2A}^{(1)}$ and $\hat{d}_{1}^{(1)}$, and
triple-cut $\hat{c}_{3m}^{(1)}$.
Reverse internal-helicity counterparts understood.
}
\label{FIG:threeminus}
\end{figure}
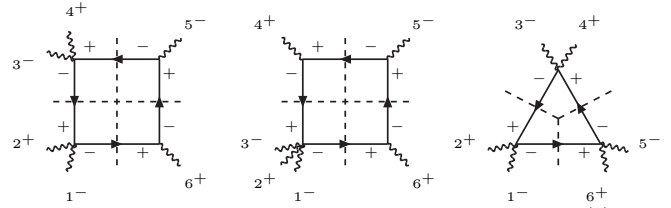


\noindent{ \bf Two-mass hard Box  $\ (1^- 2^+|3^- 4^+|5^-|6^+)$}

We define the prefactor
\bea
r_{2A}= && \nonumber \\ 
&& \hspace*{-1.5cm}
 8 {  \ 
s_{345} 
s_{56}^2
\over
\spa 2.6
\spb 3.5
\spab 6.P_{12}.3
\spab 2.P_{34}.5
\spaa{6}.{ P_{34}}.{ P_{12}}.{ 6}
\spbb{5}.{ P_{34}}.{ P_{12}}.{ 5} 
} ,
\eea
thus the result of the quadruple-cut $\hat{d}_{2A}^{(1)}$ for the configuration
displayed in Fig.\ref{FIG:threeminus} is
\bea
\hat{d}_{2A}^{(1)} = 
r_{2A} 
s_{345}^2
\spa 6.1^2
\spb 4.5^2 \,.
\eea
By reversing the internal helicities one gets
\bea
\hat{d}_{2A}^{(2)} = 
 r_{2A}
\spab 1.P_{34}.5^2
\spab 6.P_{12}.4^2
\eea
The coefficient of $F_{2A}(s_{56}, s_{345}, s_{12}, s_{34})$, 
the finite part of the two-mass-hard box, reads
\bea
d_{2A}(s_{56}, s_{345}, s_{12}, s_{34}) &=& 
2 \times {(- 2) \over s_{345} s_{56}} 
\times { \hat{d}_{2A}^{(1)} + \hat{d}_{2A}^{(2)} \over 2 } \,.
\label{eq:threeminusD2A}
\eea

\vspace{3mm}

\noindent{ \bf One-mass Box $ \ (1^- 2^+ 3^-|4^+|5^-|6^+)$}

With the prefactor
\bea
r_1 = 
8 {
s_{45} s_{56} s_{456} 
\over
\spa 4.6^2
\spab 4.P_{123}.1
\spab 4.P_{123}.3
\spab 6.P_{123}.1
\spab 6.P_{123}.3
} \ ,
\eea
the result of the quadruple-cut $\hat{d}_{1}^{(1)}$ for the configuration
displayed in Fig.\ref{FIG:threeminus} is
\bea
\hat{d}_1^{(1)} = 
r_1 
\spa 4.5^2 
\spab 6.P_{123}.2^2 \,.
\eea
By reversing the internal helicities one gets
\bea
\hat{d}_1^{(2)} = 
r_1
\spa 5.6^2 \spab 4.P_{123}.2^2 \,.
\eea
The coefficient of $F_1(s_{45}, s_{56}, s_{123})$, the finite part of the one-mass box, reads
\bea
d_1(s_{45}, s_{56}, s_{123}) &=& 
2 \times {(- 2) \over s_{45} s_{56}} 
\times { \hat{d}_1^{(1)} + \hat{d}_1^{(2)} \over 2 } \,.
\label{eq:threeminusD1}
\eea

\vspace{3mm}

\noindent{ \bf Three-mass Triangle $ \ (1^- 2^+|3^- 4^+|5^- 6^+)$}

From the triple-cut of the configuration
displayed in Fig.\ref{FIG:threeminus}, the triangle coefficient 
reads
\bea
\hat{c}_{3}^{(1)} \hspace*{-0.3cm}&=&\hspace*{-0.2cm}
 4 \  { \spa{1}.{ 2} \spab{1}.{ P_{34}}.{ 2}^2 e_1^2 
\over
  \spa{2}.{ 4} \spa{3}.{ 4} 
} \ 
{N_{e_1} \over D_{e_1}} + \Big\{e_1 \to e_2 \Big\} \,,
\label{eq:TriangleCoeff}
\eea
with
\bea
N_{e_1} \hspace*{-0.3cm}&=&\hspace*{-0.2cm}
 \Big[ 
              \!-\! \es{1}.{ 2} \spa{1}.{ 3} 
              \!+\! \Big(   (\es{1}.{ 2} \!-\! \es{3}.{ 4}) \spa{1}.{ 3} 
                  \!+\! \spaa{1}.{ P_{34}}.{ P_{12}}.{ 3}
                \Big) e_1 
 \Big]
\nonumber \\ && \hspace*{-0.8cm} \times 
  \Big[
           \es{1}.{ 2} \es{3}.{ 4} \spa{1}.{ 5} e_1 
         \!+\! \spaa{1}.{ P_{12}}.{ P_{34}}.{ 5} (-\es{1}.{ 2} \!+\! (\es{1}.{ 2} \!-\! \es{3}.{ 4}) e_1)
  \Big]^2
\nonumber \\ && \hspace*{-0.3cm} \times 
    \Big[
            \!-\! \es{1}.{ 2} \spa{1}.{ 3} \spa{2}.{ 4} 
            \!+\! \Big( 
                  (\es{1}.{ 2} \!-\! \es{3}.{ 4}) \spa{1}.{ 3} \spa{2}.{ 4} 
                \!+\! \nonumber \\ && \quad
                \!+\! \spa{3}.{ 4} \spaa{1}.{ P_{34}}.{ P_{12}}.{ 2} 
                \!-\! \spa{3}.{ 2} \spaa{1}.{ P_{34}}.{ P_{12}}.{ 4}
              \Big) e_1
    \Big] \,,
\label{eq:Nofe1}
\eea
\bea
D_{e_1} \hspace*{-0.3cm}&=&\hspace*{-0.2cm}
{  \Big[\!-\! \es{1}.{ 2} \spa{1}.{ 2} 
        \!+\! \Big(
                  (\es{1}.{ 2} \!-\! \es{3}.{ 4}) \spa{1}.{ 2} 
                \!+\! \spaa{1}.{ P_{34}}.{ P_{12}}.{ 2}
          \Big) e_1
   \Big]
}
\nonumber \\ && \hspace*{-0.8cm} \times 
{\Big[
         \es{1}.{ 2} \es{3}.{ 4} \spa{1}.{ 6} e_1 
       \!+\! \spaa{1}.{ P_{12}}.{ P_{34}}.{ 6} \Big(-\es{1}.{ 2} \!+\! (\es{1}.{ 2} \!-\! \es{3}.{ 4}) e_1 \Big)
\Big]
}
\nonumber \\ && \hspace*{-0.3cm} \times 
{\Big[ \!-\! \es{1}.{ 2} \spa{1}.{ 4} 
       \!+\! \Big((\es{1}.{ 2} \!-\! \es{3}.{ 4}) \spa{1}.{ 4} \!+\! \spaa{1}.{ P_{34}}.{ P_{12}}.{ 4} \Big) e_1
\Big] 
}
\nonumber \\ && \hspace*{-0.3cm} \times 
{ \Big[
         \!-\! \es{1}.{ 2} \spa{1}.{ 6}  
         \!+\! \Big( (\es{1}.{ 2} \!-\! \es{3}.{ 4}) \spa{1}.{ 6} \!+\! \spaa{1}.{ P_{34}}.{ P_{12}}.{ 6} \Big) e_1
 \Big] 
}
\nonumber \\ && \hspace*{-0.3cm} \times 
{ \Big[    \es{1}.{ 2} \spab{1}.{ P_{34}}.{ 3} e_1 
         \!+\! \spab{1}.{ P_{12}}.{ 3} \Big(-\es{1}.{ 2} \!+\! (\es{1}.{ 2} \!-\! \es{3}.{ 4}) e_1 \Big)
\Big]
}
\nonumber \\ && \hspace*{-0.3cm} \times 
{  \Big[ -\es{1}.{ 2} \!+\! \Big(\es{1}.{ 2} \!-\! \es{3}.{ 4} \!+\! \spab{1}.{ P_{34}}.{ 1}\Big) e_1 
  \Big] \,,
}
\label{eq:Dofe1}
\eea
where
\bea
e_{1,2} = {\es{1}.{ 2} \over 2} \ 
{ 3 \es{3}.{ 4} \!-\! \es{5}.{ 6} -\es{1}.{ 2} 
\pm \sqrt{\Delta_{12,34,56}}
\over
    \es{1}.{ 2} (\es{3}.{ 4} \!-\! \es{5}.{ 6}) 
  \!+\! \es{3}.{ 4} (-2 \es{3}.{ 4} \!+\! \es{5}.{ 6})
}\;,
\eea
with
\bea
\Delta_{12,34,56} = 
\es{1}.{ 2}^2 \!+\! \es{3}.{ 4}^2 \!+\! \es{5}.{ 6}^2 
\!-\! 2 \es{1}.{ 2} \es{3}.{ 4} 
\!-\! 2 \es{1}.{ 2} \es{5}.{ 6}
\!-\! 2 \es{3}.{ 4} \es{5}.{ 6} .
\eea
The contribution coming from reversing the inner helicities, $\hat{c}_{3}^{(2)}$,
amounts to the same value,
$
\hat{c}_{3}^{(2)} = \hat{c}_{3}^{(1)} \ ,
$
therefore the coefficient of $I_3(s_{12},s_{34},s_{56})$,
the three-mass triangle within 
the amplitude is,
\bea
{c}_{3}(s_{12},s_{34},s_{56}) &=& 2 
\times (\hat{c}_{3}^{(1)} + \hat{c}_{3}^{(2)}) \ ,
\label{eq:threeminusC3}
\eea
where the factor 2 accounts for the contribution coming from 
a fermion looping in the opposite-direction.
Although not manifest in Eqs.(\ref{eq:Nofe1},\ref{eq:Dofe1}), 
one can see analytically that in (\ref{eq:TriangleCoeff})
the dependence on $\sqrt{\Delta}$ drops out, 
because $e_1$ and $e_2$ only differ in the sign of  $\sqrt{\Delta}$. 
Therefore  $c_3$ 
is a rational function of spinor products.

\section{Form factor approach}

In the Feynman-diagrammatic approach the six-photon amplitude 
is represented by 120 one-loop diagrams which differ only 
by permutations of the external photons.

As the corresponding integrals are IR/UV finite, the Dirac algebra can be performed 
in $D=4$ dimensions. However, using algebraic tensor reduction, one has to 
work with a $(4-2\epsilon)$-dimensional loop momentum, since 
at intermediate steps scalar integrals are generated which are formally
divergent in $D=4$ dimensions. The respective coefficients will drop out in the end, 
which serves as a check of the computation.
We use the spinor helicity method 
and define projectors on 
the helicity amplitudes $A_6(--++++)$ and $A_6(-+-+-+)$ 
in such a way that -- by choosing convenient reference momenta
for the polarisation vectors -- we obtain global
spinorial factors for each amplitude. 
The resulting expressions  contain integrals with 
scalar products of external vectors and loop momenta in the numerator.
For the six-point integrals all scalar products between the loop momentum
and external momenta are reducible, i.e.\ they can  be written as differences of 
inverse propagators. As a result, at most rank-one six-point functions have
to be evaluated. For the $(N<6)$-point functions at most three loop
momenta remain in the numerator. As was shown in \cite{binoth2},
this can be related to the cut-constructibility argument of \cite{Bern:1994zx}.
To reduce the  irreducible tensor integrals to scalar integrals 
we use the algebraic approach
outlined in \cite{method1}, 
which leads to a representation of the amplitude in terms of 
the basis functions $I_{3}, F_{1},F_{2A},F_{2B}$.  
The formalism 
is implemented using {\tt FORM} \cite{FORM}.
The coefficients of the basis functions 
are then stored and simplified 
further with {\tt Maple} and/or {\tt Mathematica}. 
The same setup was already used for
simpler 4- and 5-point loop amplitudes~\cite{golem_apps}.

Although  
only a restricted set of functions actually needs to be evaluated  
due to the symmetry properties of the amplitude,  
the symmetry relations serve as a stringent check on the
implementation. 
Therefore, and in order to 
test our setup in view of future applications,  
all 120 diagrams have been calculated.

We stress that the resulting expressions for the coefficients, 
although they are  not evaluated in terms of spinor products, 
allow for a fast numerical evaluation. We note that
in all expressions at most one power of inverse
Gram determinants survives, which is intrinsic to the chosen
function basis. 

\begin{figure}[h]\label{fig_amp}
\vspace*{-1.0cm}
\includegraphics[width=9.0cm]{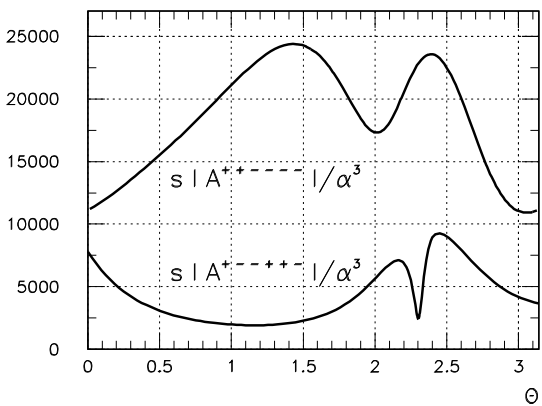}
\vspace*{-2.0cm}
\caption{The modulus of the normalised six-photon amplitudes
$s\,|{A}_6(++----)|/\alpha^3$ and 
$s\,|{A}_6(+--++-)|/\alpha^3$ plotted
for the kinematics  as defined in \cite{nagy}.
}
\end{figure}

We have cross-checked the result obtained by the 
form factor approach  with the one achieved by u\-sing 
the 
cutting rules and find perfect agreement. 
We have further  compared to the recent numerical result of Nagy and Soper~\cite{nagy}:
mapping our helicity configuration onto their one, according to 
$
A_6(2^-,1^+,3^-,4^+,6^-,5^+)=A_6(1^+,2^-,3^-,4^+,5^+,6^-) \,,
$
and using the same kinematics as in Fig.\,5 of \cite{nagy}, we find the
result shown in Fig. \ref{fig_amp}, which agrees with Nagy and Soper within
their plot range.
We note that our results are produced by simply evaluating  
the analytical expressions obtained by the form factor approach, 
  therefore we do not have  numerical errors. The peak structures in the plots stem from complicated
phase patterns in the amplitudes.

\section{Conclusions}
The six-photon amplitudes can be fully expressed as linear combinations of 
known one-loop master integrals with three and four external momenta. 
Using the techniques of form factor decomposition~\cite{method1}
and of multiple 
cuts~\cite{Britto:2004nc,Britto:2005ha,Britto:2006sj,Mastrolia:2006ki},
we have derived analytic expressions for the coefficients of the master 
integrals. The expressions obtained from the multiple-cut method are typically 
more compact than the form factor results, such that we decided to quote only 
the former. All coefficients agree numerically. We fully confirm earlier 
purely numerical results~\cite{nagy} for the six-photon amplitudes. 
Our calculation demonstrates the applicability of the form factor decomposition 
and multiple-cut methods to non-trivial multi-leg processes at one-loop, 
and illustrates that both methods can be formulated in the same integral
basis. 
In future applications, one could therefore envisage to combine both
methods. Both approaches used here can also be applied to amplitudes 
containing massive particles. 
Our result shows that the analytical evaluation
of one-loop amplitudes with similar kinematics
relevant for the LHC is feasible.

\vspace{3mm}

{\bf Note added:} Shortly after the appearance of this Letter, 
numerical results on the same subject 
were presented by Ossola, Papadopoulos and Pittau using a reduction
method at the integrand level~\cite{opp}. Full consistency was found 
where applicable. Moreover, the triangle coefficient (\ref{eq:TriangleCoeff}) in 
$A_6(-+-+-+)$ was rederived recently by Forde~\cite{Forde:2007mi} 
using an independent method. 
Full agreement was found with our result.

\vspace{3mm}

\noindent{\bf Acknowledgements} 

We would like to thank Jean-Philippe Guillet, Christian Schubert 
and  Daniel Egli for collaboration during early stages;
Simon Badger, Ruth Britto and Andrea Ferroglia, for clarifying discussions.
GH would like to thank the ITP at Z\"urich  University 
for its hospitality.
This research was supported in part by Marie-Curie-EIF under contract 
MEIF-CT-2006-024178, by 
the Swiss National Science Foundation
(SNF) under contract 200020-109162,  
the UK Particle Physics and Astronomy  Research Council (PPARC) 
and the Scottish Universities Physics Alliance (SUPA).


\end{document}